\def\gd{Gd$^{3+}$}
\def\nd{Nd$^{3+}$}

\def\eu{Eu$^{3+}$}
\def\co{CuO$_2$}
\def\lco{$\rm La_2CuO_4$}
\def\lsco{$\rm La_{2-x}Sr_xCuO_4$}

\def\lesco{$\rm La_{2-x-y}\-Eu_y\-Sr_x\-CuO_4$}
\def\lgesco{$\rm La_{1.99-x-y}Sr_xGd_{0.01}Eu_yCuO_4$}
\def\lnsco{$\rm La_{2-x-y}\-Nd_y\-Sr_x\-CuO_4$}

\def\dh{$\rm \Delta H$}
\def\dht{$\rm \Delta H(T)$}
\def\gcu{g\raisebox{-1.0ex}{\scriptsize Cu}}

\documentstyle[preprint,aps]{revtex}

\begin{document}
\draft
\preprint{}
\title{\raisebox{1cm}{\rightline{\it to appear in
Phys.Rev.B (Rapid Com.), Feb., 97}}
Slow antiferromagnetic dynamics  in the low
temperature tetragonal phase of $\rm\bf La_{2-x}Sr_xCuO_4$ as
                        revealed by ESR
                       of Gd spin probes}

\author{V.Kataev and B.Rameev}
\address{Kazan Institute for Technical Physics, Russian Academy
of Sciences, 420029 Kazan, Russia}
\author{B.B\"uchner, M.H\"ucker and R.Borowski}
\address{II. Physikalisches Institut, Universit\"at zu K\"oln,
Z\"ulpicher Str. 77, 50937 K\"oln, Germany}

\date{November 12, 1996}

\maketitle

\begin{abstract}
Measuring the ESR of Gd spin probes we have studied the magnetic
properties of the copper oxide planes in the low temperature
tetragonal  (LTT) phase of Eu doped \lsco. The data give
evidence that at particular levels of Sr and Eu doping the
frequency of the antiferromagnetic fluctuations $\rm\omega_{sf}$
in the LTT phase dramatically decreases at low temperatures by
almost three orders of magnitude. However, no static magnetic
order has been found for $\rm T\geq 8\,K$ in contrast to the
observation by neutron scattering of stripe ordering of spins
below 50\,K in a Nd doped \lsco\ single crystal. To our opinion
static order in the Nd doped compound is induced due to the
interaction between the Cu spins with the rare earth magnetic
moments. Therefore, a really characteristic property of the
magnetism in the LTT structural phase may be not static magnetic
order at elevated temperatures but rather extremely slow
antiferromagnetic dynamics.
\end{abstract}

\pacs{\mbox{PACS: 74.25.Ha; 74.72.Dn; 76.30.Kg}
\centerline{%
\raisebox{-4cm}{%
\fbox{%
\parbox{9cm}{\it
If you have read this paper and wish to be included in a mailing
list that we maintain on the subject, please send e-mail to:
{\bf kataev@colorix.ph2.uni-koeln.de}}}}}
}  

\narrowtext

The structural phase transition in \lsco\  from the low
temperature orthorhombic (LTO) phase to the low temperature
tetragonal (LTT) phase \cite{Buchner,Craw} has recently
attracted much attention. Neutron diffraction experiments on a
non superconducting $\rm La_{1.48}\-Nd_{0.4}\-Sr_{0.12}\-CuO_4$
single crystal show an unusual type of magnetic order below
50\,K in the form of antiferromagnetic (AF) domains ('stripes')
separated by walls in which holes are segregated \cite{Tranq}.
It was argued that such stripe order is feasibly a result of
pinning by particular direction of tilting of Cu--O octahedra in
the LTT phase of dynamically correlated AF regions found earlier
in superconducting \lsco\ \cite{Cheong}. Furthermore, this
interpretation has been connected with an idea of a frustrated
phase separation (see e.g. \cite{Emery}) pointing at the
importance of this observation for clarification of the nature
of high temperature superconductivity (HTSC)
\cite{Emery,Zaanen}.

In this Communication we present the first results of the
systematic study of the magnetic properties of the LTT phase of
\lesco\ by means of ESR. As a spin probe in the ESR experiments
a \gd\ ion which substitutes the rare earth (RE) site in the
structure has been chosen. A simple qualitative analysis of the
observed temperature dependence of the \gd\ ESR linewidth gives
evidence that at certain concentrations of dopants the frequency
of spin fluctuations in the \co\ planes with the LTT structure
slows with lowering the temperature down to nearly $\rm
10^{10}\, sec^{-1}$. However, within the temperature range of
study ($\rm 8\leq T \leq 300\,K$) no signatures of a really
static AF order have been observed in the samples investigated
so far. This implies that in a frame of the stripe model spin
correlations remain dynamic even in the LTT phase.

The polycrystalline samples of \lsco\ in which part of La ions
was substituted by \eu\ (up to 12\% relative to La) were
prepared and characterized as described elsewhere \cite{Breuer}.
\gd\ ions were added in amount of 1\%. The role of Eu is to
induce the transition to the LTT phase due to mismatch in ionic
radii while hole concentration is tuned independently by Sr
doping $(\rm 0.05\leq x \leq 0.20)$ \cite{Buchner1}. In this
respect the small percentage of \gd\ does not affect the
relevant physical properties. Substitution by Eu instead of Nd
was chosen mainly because \eu\ in its ground state possesses
only Van--Vleck paramagnetism and the influence on bulk
susceptibility and ESR of thermally excited magnetic states
lying 400\,K above the ground state is much weaker in comparison
to magnetic \nd\ ions and can be correctly subtracted
\cite{Hucker}. Moreover, in the case of Eu substitution there is
no influence of permanent magnetic moments at the RE sites on
the magnetism of the \co\ planes.

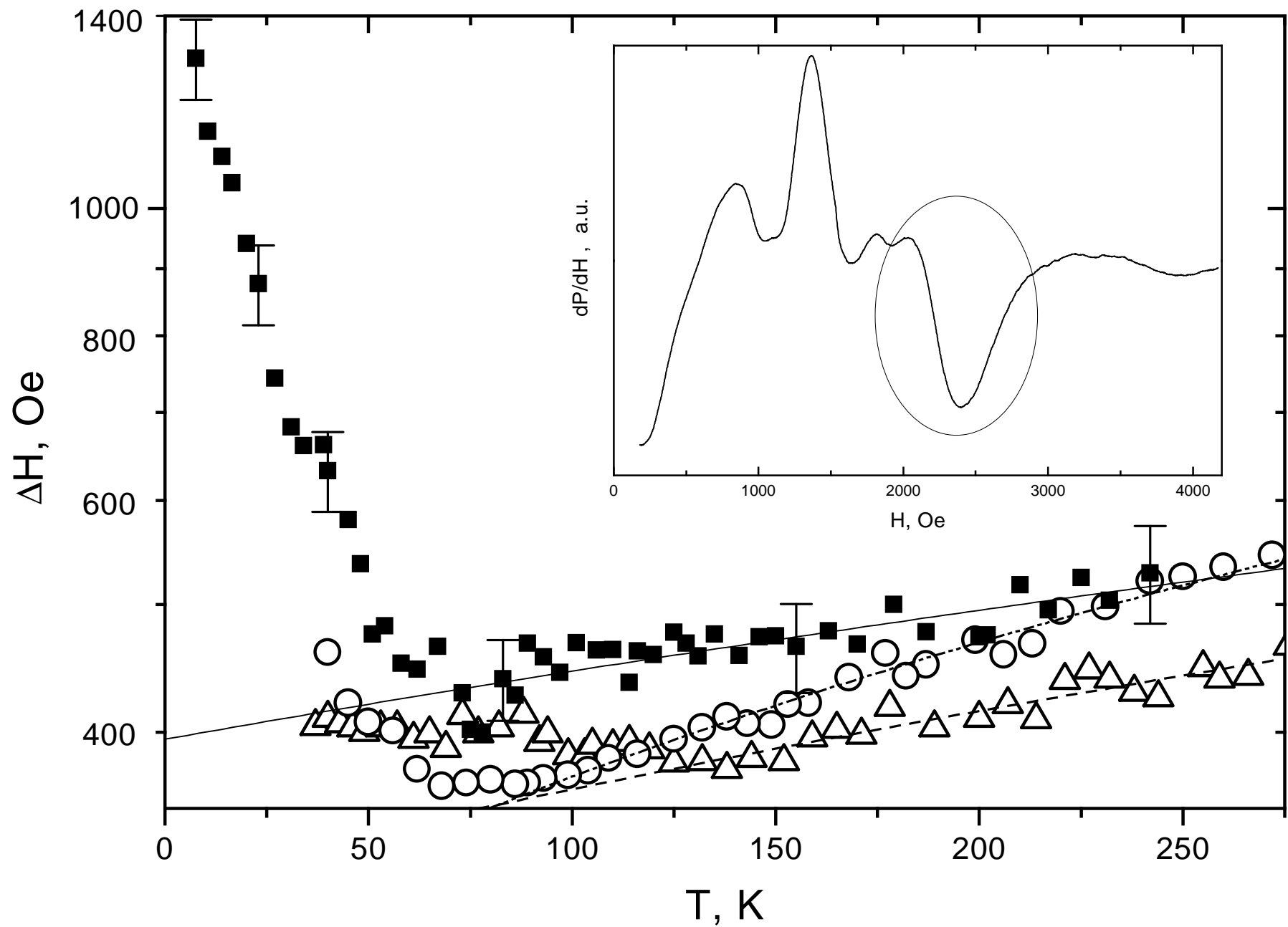
\begin{figure}
\caption{Temperature dependence of the width of the central
component of the \gd\ ESR spectrum of \lgesco\  (symbols)
together with corresponding linear fit $\rm a+bT$ of its high
temperature part (straight lines): $\rm x=0.10,\ y=0.0$
($\triangle$), dash line $\rm 309+0.53T$; $\rm x=0.17,\ y=0.15$
($\bigcirc$), dash--dot line $\rm 272+0.98T$; $\rm x=0.10,\
y=0.24$ (\protect\rule{.5em}{.5em}),
solid line $\rm 395+0.5T$. For the
latter sample the contribution due to thermally excited magnetic
states of \eu\ lying above 400\,K has been subtracted so that to
keep the slope b the same as for the samples with lower Eu
dopings. In inset the fine structure split ESR spectrum of \gd\
is shown and the component which \dht\ dependence is plotted in
the main figure is encircled.}
\label{dh-t}
\end{figure}

\gd\ ESR spectra of \lgesco\ measured at a frequency of 9.3\,GHz
show for all studied samples a fine structure due to the small
splitting of the ground state multiplet $\rm ^8S_{7/2}$ of a
\gd\ ion in the crystalline electrical field (see inset in
Fig.\ref{dh-t}). The analysis of the spectra has been developed
previously \cite{Kat_1,Kat_2}. Typical temperature dependences
of the width \dh\ of the central component of the spectrum
(encircled in inset of Fig.\ref{dh-t}) are shown in
Fig.\ref{dh-t}. For $\rm T>80-100\,K$ \dh\ increases linearly
with temperature as $\rm a+bT$. In agreement with our earlier
findings \cite{Kat_2} the slope $\rm b=d(\Delta H)/dT$ increases
with increasing the Sr (i.e. hole) concentration. At a fixed Sr
content b does not change significantly until the Eu
concentration becomes higher than $\rm\sim 8\%$. For these Eu
contents \dh\ increases due to relaxation via thermally excited
magnetic states of \eu\ \cite{Zysler} which is appreciable for
$\rm T>100\,K$.

The remarkable feature of \dht\ is the qualitative change of its
behavior in the low temperature region where a pronounced
broadening of the \gd\ ESR spectrum is observed (see
Fig.\ref{dh-t}).  This effect was found to depend on both Sr and
Eu content. In particular, at a fixed Sr concentration x the
broadening increases with increasing the Eu concentration y. The
details of the doping dependence will be published separately
\cite{Kat_3}. Hereafter we focus on the low temperature \dht\
dependence of a single representative sample with the
composition $\rm La_{1.65}Sr_{0.1}Gd_{0.01}Eu_{0.24}CuO_4$, i.e.
$\rm x=0.10,\ y=0.24$. This compound is found to be in the non
superconducting LTT phase below $\rm\sim 130\,K$
\cite{Buchner1}. For this sample the broadening of the \gd\ ESR
linewidth is most pronounced and the observation of the ESR
spectrum is not obscured by the large field dependent drift of
the base line due to superconductivity.

As we have shown earlier \cite{Kat_2}, the linear temperature
dependence of the \gd\ ESR linewidth in the normal state of the
LTO phase of \lsco\ is a result of the Korringa relaxation of Gd
spins  due to a small but finite exchange coupling of \gd\ ions
to the mobile holes in the \co\ planes. As the Gd ion probes the
spin dynamics of the copper oxide planes, a very pronounced
deviation of \dht\ from linearity in the low temperature region
observed in the LTT phase for the Eu doped sample is obviously
due to strong changes of the spectrum of spin excitations in
these key  elements of the structure of HTSC compounds. At this
point one should mention that a small deviation from the linear
\dht\ dependence is noticeable even for Eu free sample (see
Fig.\ref{dh-t}) which is nominally in the LTO phase. To our
opinion, this is due to the fact that Gd spin probe itself
creates an LTT distortion of the lattice in the nearest
surrounding and therefore may locally modify the spin dynamics
in the \co\ plane.

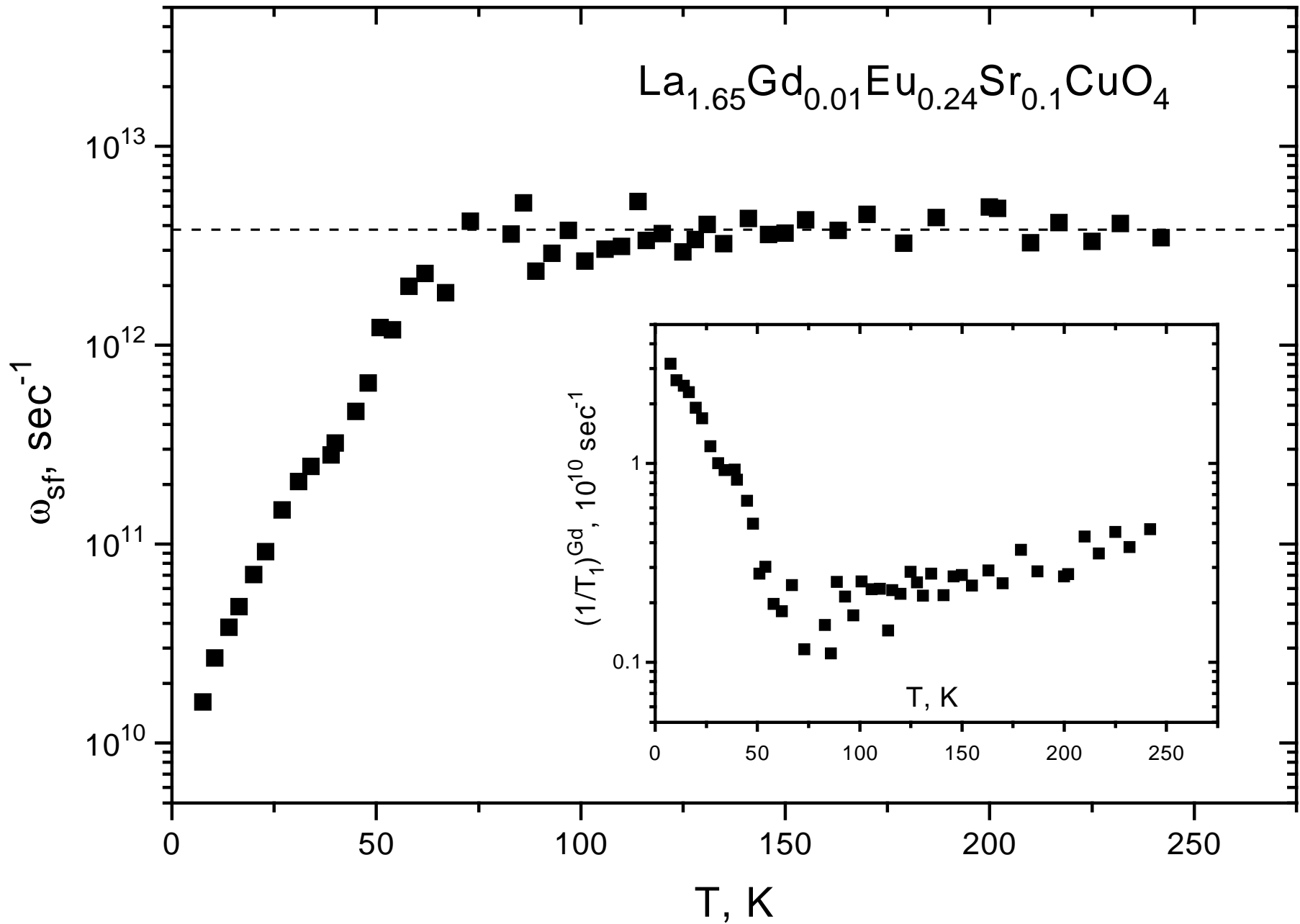
\begin{figure}
\caption{Temperature dependence of the spin fluctuation
frequency for the $\rm
La_{1.65}\-Sr_{0.1}\-Gd_{0.01}\-Eu_{0.24}\-CuO_4$
sample estimated using (\protect\ref{wsf}). Inset: the Gd spin
relaxation as a function of T extracted from the measured
linewidth according to (\protect\ref{Gdrelax}).}
\label{omegasf}
\end{figure}

To evaluate the influence of the Cu spin dynamics on spin
relaxation of \gd\ ions we first recall that the measured ESR
linewidth of magnetic ions in metals is usually separated into
two parts \cite{Barnes}:
\begin{equation}
\label{dHtotal}
\rm
\Delta H = \left(\Delta H\right)_0 + \left(\Delta H\right)_{relax}.
\end{equation}
$\rm \left(\Delta H\right)_0$ is the so called residual width
which arises due to a number of static reasons such as
inhomogeneities of the crystal field potential and local
magnetic fields, hyperfine coupling, anisotropic spin--spin
interactions, etc. The second, homogeneous, contribution $\rm
\left(\Delta H\right)_{relax}$, which is most important in our
analysis of \dht,  is determined by the spin relaxation of an
ion. For the case of the fine structure split ESR spectrum of a
\gd\ ion the relaxation determined part of the width of an
individual component is related to the spin relaxation rate $\rm
1/T_1$ as $\rm \left(\Delta H\right)_{relax}=M^2(1/\gamma T_1)$,
where $\gamma$ is the gyromagnetic ratio and M is the matrix
element of the corresponding Zeeman transition \cite{Barnes}.
Computer modeling of the ESR spectra of \gd\ in \lsco\ gives a
value $\rm M^2\approx 0.5$ for the component of our interest.
Thus, the Gd spin relaxation rate reads:
\begin{equation}
\label{Gdrelax}
\rm
\left(\frac{1}{T_1}\right)^{Gd}=\frac{\gamma}{M^2}
\left[\Delta H - \left(\Delta H\right)_{0}\right] .
\end{equation}
Assuming $\rm \left(\Delta H\right)_0$ to be equal to the
parameter {\it a} of the high temperature linear fit $\rm {\it
a} + {\it b}T$ of the \dht\ dependence (see Fig.\ref{dh-t}), we
plot in inset of Fig.\ref{omegasf} the dependence of
$\rm\left(1/T_1 \right)^{Gd}$ versus temperature for the sample
with $\rm x=0.10, y=0.24$.

Similar to the general expression of the nuclear relaxation rate
\cite{Moriya}, the spin relaxation of the \gd\ ions due to their
coupling with the \co\ planes can be written in terms of the
dynamic susceptibility of the planes
$\rm\chi\raisebox{0.5ex}{$\prime\prime$}({\bf q},\omega)$ as:
\begin{equation}
\label{relax-esr}
\rm
\frac{1}{T_1}=\frac{kT\cdot J^2 _{Gd-Cu}}{(\gcu\mu _B\hbar)^2}
\lim_{\omega\to 0}\sum_{\bf q}^{} f({\bf q})
\frac{\chi\raisebox{0.5ex}{$\prime\prime$}({\bf q},\omega)}{\omega} .
\end{equation}
Here $\rm J _{Gd-Cu}$ is a coupling constant which determines
the strength of the exchange interaction between Gd and Cu
spins, $\rm \gcu$ is the g--factor of Cu, and $\rm f({\bf q})$
is a geometrical form factor. However,
$\rm\chi\raisebox{0.5ex}{$\prime\prime$}({\bf q},\omega)$ in the
layered cuprates is described differently in various theoretical
approaches. This makes a quantitative interpretation of the data
model dependent as will be discussed in a separate publication
\cite{Kat_3}. Fortunately, as we shall see below, a considerable
qualitative insight into the low frequency spin dynamics in Eu
doped \lsco\ can be already provided if we rewrite
Eq.\ref{relax-esr} in a simplified form:
\begin{equation}
\label{relax-wsf}
\rm
\frac{1}{T_1}\sim\frac{k\cdot J^2 _{Gd-Cu}}{(\gcu\mu
_B\hbar)^2}\chi _0 \cdot \frac{T}{\omega_{sf}},
\end{equation}
where $\chi _0$ is the measured static susceptibility and
$\rm\omega_{sf}$ is the frequency of the spin fluctuations in
the \co\ plane. Then, combining (\ref{Gdrelax}) and
(\ref{relax-wsf}) we obtain:
\begin{equation}
\label{wsf}
\rm
\omega_{sf}\sim
\frac{k\cdot J^2 _{Gd-Cu}M^2\chi_0}{(\gcu\mu _B\hbar)^2\gamma}
\frac{T}{\Delta H - \left(\Delta H\right)_{0}}.
\end{equation}

To extract the values of $\rm \omega_{sf}$ from the experimental
data using (\ref{wsf}) one has to know $\rm J_{Gd-Cu}$ and
$\chi_0$. An estimate of the strength of the rare earth Cu
exchange coupling in the hole doped lanthanum copper oxide can
be obtained from recent specific heat measurements of \lnsco\
which show for samples with suppressed superconductivity in the
LTT phase a Schottky anomaly at low temperatures \cite{Nguyen}.
It can be attributed to the splitting of the ground state
Kramers doublet of \nd\ due to slowly fluctuating or even static
magnetic field $\rm H_{int}\sim J_{RE-Cu}<\!\mu _{Cu}\!>$ of the
order of 1\,T transferred from the Cu spin lattice. Similar low
temperature specific heat has also been observed in Gd doped
\lsco\ samples with the LTT structure \cite{Czjzeck}. Although
in this case the interpretation of the Schottky anomaly is more
complicated due to the crystalline field fine structure
splitting of the ground state multiplet of the \gd\ ion, the
estimate gives a similar value of the transferred magnetic field
at the RE site \cite{Rettori}. Taking the value of $\rm <\!\mu
_{Cu}\!>\simeq 0.5\mu _B$ we obtain $\rm J_{RE-Cu}\sim 5\,K$.

Measurements of the static susceptibility on samples with
similar stoichiometry  but without Gd show that $\chi _0$
changes with temperature not more than within a factor of two.
Therefore, for the following estimates it can be taken as
constant with a value $\rm\chi _0\approx
2\cdot10^{-4}\,emu/mole$.

With these values of the exchange constant and static
susceptibility we plot in Fig.\ref{omegasf} the temperature
dependence of the spin fluctuation frequency evaluated from the
experimental data according to (\ref{wsf}). As it can be seen
from this Figure, $\rm\omega_{sf}$ is temperature independent
above $\rm\sim 75\,K$. Although the obtained energy scale of
these fluctuations  $\rm \hbar\omega_{sf}\approx 40\,K$ is
consistent with that probed in the NMR experiments on \lsco\
(see e.g., the analysis in Ref.\cite{MMP2}), its value is, of
course, sensitive to the choice of parameters in (\ref{wsf}).
However, most important is that the temperature dependence of
$\rm\omega_{sf}$ presented in Fig.\ref{omegasf} demonstrates a
qualitative change of the spin dynamics in the LTT phase at low
temperatures. A steep decrease of the fluctuation frequency
below $\rm\sim 70\,K$ by more than two orders of magnitude
points at a dramatic slowing of spin fluctuations in the \co\
planes with the LTT structure. Such slowing of spin dynamics has
a profound effect on the spin relaxation of Gd ions leading to
strong enhancement of $\rm \left(1/T_1\right)^{Gd}$ and,
consequently, to the experimentally observed broadening of the
\gd\ ESR line. A similar pronounced increase of $\rm (1/T_1)$ of
Cu nuclei has been found for insulating \lco\ at temperatures
approaching the Neel transition temperature $\rm T_N$ (see e.g.,
Ref.\cite{Imai}). Although in the case of Eu doped \lsco\ a
strong tendency of the system to long range order at $\rm T_N>0$
at finite levels of hole doping is evident from the data, a
really static AF order is not found within the temperature range
of study. It would manifest itself in a narrowing and splitting
of the \gd\ ESR spectrum. Instead we observe in the LTT phase a
very slow spin dynamics of AF correlated regions. A quantitative
evaluation of the spatial extent of these correlations (i.e. the
AF correlation length $\xi$) from the ESR data requires a
particular model of spin relaxation in the cuprates. We defer
such analysis to a further publication \cite{Kat_3}. Here we
only mention that the phenomenological model of the nuclear spin
lattice relaxation developed by Millis, Monien and Pines (MMP)
\cite{MMP} seems to be applicable in our case \cite{commen}. In
the frames of this model we arrive at the result that the
correlation length $\xi$ in the hole doped \co\ planes with the
LTT structure increases up to more that 100 lattice constants
\cite{Kat_3}. This should correspond to a dramatic decrease of
the fluctuation frequency down to $\rm\sim 10^{10}\,sec^{-1}$
which matches well with our above made qualitative estimates
(see Fig.\ref{omegasf}).

However, we emphasize that independent on a particular
theoretical model the mere fact of the strong enhancement of the
\gd\ spin relaxation upon lowering T shows a rapid slowing down
of AF fluctuations and the absence of long range AF order at
elevated temperatures. This is in contrast to the results of
Tranquada et al. \cite{Tranq} whose neutron scattering data show
static ordering of spins and charges in a $\rm
La_{1.48}\-Nd_{0.4}\-Sr_{0.12}\-CuO_4$ single crystal already
below 50\,K. This contradiction could be due to different
characteristic energies of neutron diffraction and ESR
experiments, respectively. The stripe correlations may be
already static from the point of view of neutron scattering but,
in fact, remain dynamic with a very slowly fluctuation frequency
$\rm\omega_{sf}\sim 10^{10} - 10^{11}\, sec^{-1}$ from the point
of view of ESR. However, in $\mu$SR experiments with
characteristic frequency $\rm\sim 10^{6}\, sec^{-1}$ magnetic
order in $\rm La_{1.85-x}\-Nd_{x}\-Sr_{0.15}\-CuO_4$ has also
been observed at rather high $\rm T\sim 28\,K$ \cite{Wagener1}.
In contrast to this, first $\mu$SR results on the Eu doped
system show no signs of magnetic order at elevated temperatures
\cite{Wagener2}. Hence, static magnetic order occurs possibly
due to interaction of the Cu spin system with magnetic moments
of Nd which obviously is not the case for Eu doped \lsco.
Therefore, from our data we conclude  that the main feature of
magnetism of the LTT structural phase of the lanthanum strontium
copper oxide and its possible relation to HTSC is not AF order
but rather a dramatic slowing down of spin dynamics.

The authors would like to thank G.Khaliullin, W.Brenig and
G.Teitel'baum for helpful discussions. The work in Cologne was
supported by the Deutsche Forschungsgemeinschaft through SFB 341
and the work in Kazan was supported in parts by the State HTSC
Program of the Russian Ministry of Sciences (project \#940045)
and by the Russian Foundation for Basic Research under grant
\#95--02--05942. One of us (V.K.) acknowledges NATO support
under the Collaborative Research Grant HTECH.EV 960286.


\newpage

\begin{references}

\bibitem{Buchner}
B.B\"uchner, et al., Physica {\bf C 185--189}, 903 (1991).

\bibitem{Craw}
M.K.Crawford et  al.,  Phys.Rev.
{\bf B44}, 7749 (1991).

\bibitem{Tranq}
J.M.Tranquada et al.,
Nature {\bf 375}, 561 (1995);
Phys.Rev. {\bf B54}, 7489 (1996).

\bibitem{Cheong}
S.Cheong et al.,
Phys.Rev.Lett. {\bf 67}, 1791 (1991);
T.Mason et al.,
Phys.Rev.Lett. {\bf 68}, 1414 (1992);
T.Thurston et al.,
Phys.Rev. {\bf B46}, 9128 (1992).

\bibitem{Emery}
V.J.Emery and S.A.Kivelson,
Physica {\bf C235--240}, 189 (1994);
Physica {\bf C263}, 44 (1996);
U.L\"ow et al.,
Phys.Rev.Lett. {\bf 72}, 1918 (1994);
M.I.Salkola, V.J.Emery and S.A.Kivelson,
Phys.Rev.Lett. {\bf 77}, 155 (1996);

\bibitem{Zaanen}
H.Eskes et al.,
Phys.Rev. {\bf B54}, R723 (1996).


\bibitem{Breuer}
M.Breuer et al.,
Physica {\bf C208} 217 (1993).


\bibitem{Buchner1}
B.B\"uchner et al., J.Low Temp.Phys. {\bf
95}, 285 (1994);
B.B\"uchner et al.,
Physica {\bf C235--240}, 281 (1994).

\bibitem{Hucker}
M.H\"ucker, Diplomarbeit, Universit\"at zu K\"oln, 1994.

\bibitem{Kat_1}
V.E.Kataev et al.,
JETP Lett. {\bf 48}, 476 (1988);

\bibitem{Kat_2}
V.E.Kataev et al.,
JETP Lett. {\bf 56}, 385 (1992);
V.E.Kataev et al.,
Phys.Rev. {\bf B48}, 13042 (1993).

\bibitem{Zysler}
R.D.Zysler et al., Phys.Rev. {\bf B44}, 9467 (1991).

\bibitem{Kat_3}
V.Kataev et al., in preparation.

\bibitem{Barnes}
S.E.Barnes, Adv.Phys. {\bf 30}, 801 (1981).

\bibitem{Moriya}
T.Moriya,
J.Phys.Soc.Jpn. {\bf 18}, 516 (1963).


\bibitem{Nguyen}
Q.Nguyen, Diplomarbeit, Universit\"at zu K\"oln, 1995.

\bibitem{Czjzeck}
G.Czjzeck et al., unpublished.

\bibitem{Rettori}
C.Rettori et al. [Phys.Rev. {\bf B47}, 8156 (1993)] have
explained the splitting of the \gd\ ESR spectrum in the
insulating AF ordered \lco\ single crystal using the value of
$\rm H_{int}$ at the RE site of the order of 0.1\,T. This
smaller value is possibly due to the presence of interstitial
oxygen typical for samples with $\rm x=0$. We emphasize that the
value of the transferred field of the order of 1\,T yields an
upper limit of the RE--Cu exchange and thus an {\em upper} limit
of the fluctuation frequency $\rm\omega_{sf}$.

\bibitem{MMP2}
H.Monien, P.Monthoux and D.Pines,
Phys.Rev. {\bf B43}, 275 (1991).

\bibitem{Imai}
T.Imai et al.,
Phys.Rev.Letts. {\bf 70}, 1002 (1993).


\bibitem{MMP}
A.J.Millis, H.Monien and D.Pines,
Phys.Rev. {\bf B42}, 167 (1990);
H.Monien, D.Pines and M.Takigawa, {\it ibid}, {\bf 43} 258
(1991).

\bibitem{commen}
Although the assumption of the MMP model that
$\rm\chi\raisebox{0.5ex}{$\prime\prime$}({\bf q},\omega)$ is
peaked at the AF wave vector $\rm{\bf Q}(\pi/a,\pi/a)$
contradicts with the results of the neutron scattering
measurements which observe incommensurate spin correlations (see
Ref.\cite{Cheong,Tranq}), the are reasonable possibilities to
reconcile this contradiction. See e.g., a discussion by
V.Barzykin, D.Pines and D.Thelen, [Phys.Rev. {\bf B50}, 16052
(1994)] and Y.Zha, V,Barzykin and D.Pines [Phys.Rev. {\bf B54},
7561 (1996)].

\bibitem{Wagener1}
W.Wagener et al., preprint.


\bibitem{Wagener2}
W.Wagener et al., unpublished.

\end{references}
\end{document}